\documentclass[sigconf]{acmart}

\usepackage{subcaption}

\AtBeginDocument{%
  }

\graphicspath{{./Figs/}}

\copyrightyear{2026}
\acmYear{2026}
\setcopyright{cc}
\setcctype{by}
\acmConference[C\&C '26]{Creativity and Cognition}{July 13--16, 2026}{London, United Kingdom}
\acmBooktitle{Creativity and Cognition (C\&C '26), July 13--16, 2026, London, United Kingdom}
\acmDOI{10.1145/3803784.3816847}
\acmISBN{979-8-4007-2583-8/2026/07}

\begin{document}

\title{\textit{CapSenseBand}: Sustaining Cross-Disciplinary Creativity When Stitches Must Meet Signals}

\newcommand{\insta}{\textsuperscript{a}}
\newcommand{\instb}{\textsuperscript{b}}
\newcommand{\instc}{\textsuperscript{c}}
\newcommand{\instd}{\textsuperscript{d}}
\newcommand{\authmarker}[1]{\textsuperscript{{\fontsize{9}{9}\selectfont #1}}}
\newcommand{\equalcontrib}{\authmarker{*}}
\newcommand{\equalcorr}{\authmarker{\textdagger}}

\author{Sark Pangrui Xing\insta\equalcontrib}
\orcid{0000-0002-2273-4772}
\affiliation{%
  \institution{}
  \city{}
  \country{}}
\email{sark.xing@connect.polyu.hk}

\author{Hongci Hu\instb\instd\equalcontrib}
\orcid{0000-0001-7573-5229}
\affiliation{%
  \institution{}
  \city{}
  \country{}}
\email{21037958r@connect.polyu.hk}

\author{Lai Wei\instc}
\orcid{0000-0002-5476-5450}
\affiliation{%
  \institution{}
  \city{}
  \country{}}
\email{cbs-lai.wei@polyu.edu.hk}

\author{Le Fang\insta}
\orcid{0000-0003-1860-4008}
\affiliation{%
  \institution{}
  \city{}
  \country{}}
\email{leonardo-le.fang@connect.polyu.hk}

\author{Ziqian Bai\instd}
\orcid{0000-0001-9562-4315}
\affiliation{%
  \institution{}
  \city{}
  \country{}}
\email{baizq@sustech.edu.cn}

\author{Kinor Shou-xiang Jiang\instb\equalcorr}
\orcid{0000-0002-5151-9481}
\affiliation{%
  \institution{}
  \city{}
  \country{}}
\email{kinor.j@polyu.edu.hk}

\author{Stephen Jia Wang\insta\equalcorr}
\orcid{0000-0001-9835-9932}
\affiliation{%
  \institution{}
  \city{}
  \country{}}
\email{stephen.j.wang@polyu.edu.hk}

\renewcommand{\authors}{Sark Pangrui Xing, Hongci Hu, Lai Wei, Le Fang, Ziqian Bai, Kinor Shou-xiang Jiang, and Stephen Jia Wang}
\renewcommand{\shortauthors}{Xing et al.}

\begin{abstract}
  Wearable sensing systems increasingly depend on textiles that are both materially wearable and electronically functional. Their design requires collaboration between textile designers, who reason through stitches, yarn behavior, and machine constraints, and interaction designers, who reason through electrodes, signal paths, and insulation. However, these forms of expertise do not easily translate across disciplinary boundaries. This poster presents \textit{CapSenseBand}, a knitted capacitive-sensing wristband developed through a research-through-design process organized around Analysis, Synthesis, and Detailing. We document an artifact chain spanning material swatches, a rapid wearable prototype, Paper Models as shared negotiation surfaces, a double-layer knitted structure, and an insulated Swept Frequency Capacitive Sensing breakout board. We show how Paper Models functioned as boundary objects, helping collaborators externalize intent, negotiate spatial and technical constraints, and preserve disciplinary expertise while converging on a shared design. We contribute a reusable swatch-to-sleeve pattern for material-centered HCI: keep discipline-specific probes open early, then converge through artifacts that make material, spatial, and electronic decisions legible before fabrication locks them in.
\end{abstract}



\begin{teaserfigure}
  \includegraphics[width=\textwidth]{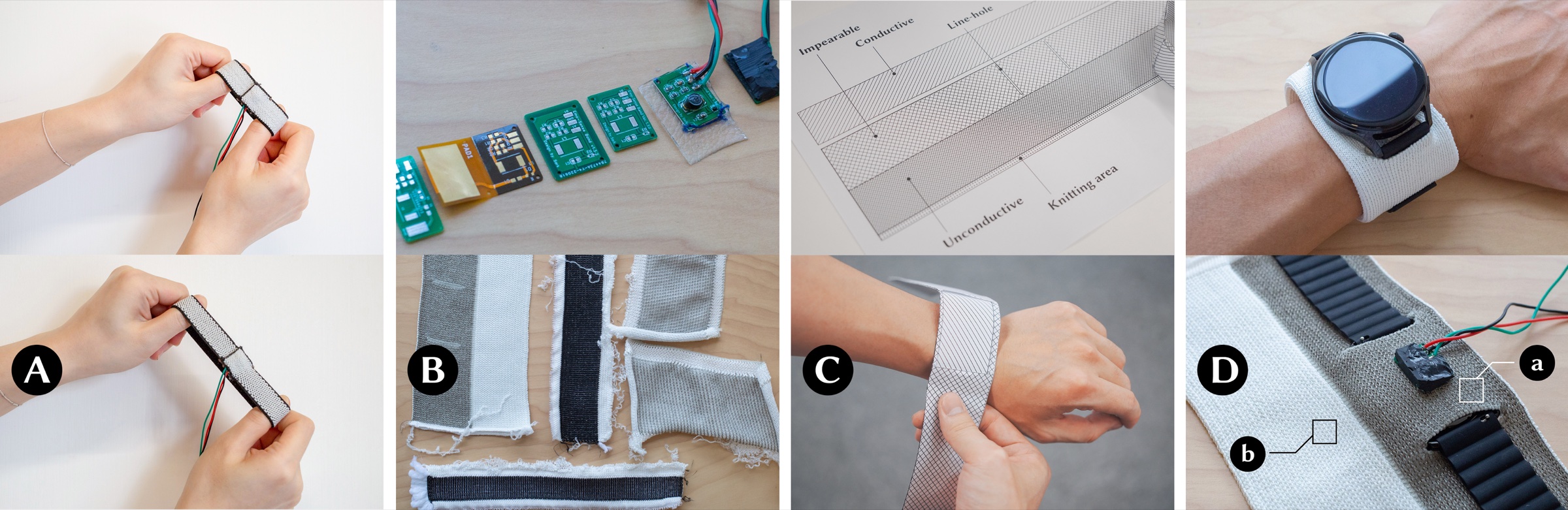}
  \caption{Evolution of \textit{CapSenseBand} through interdisciplinary work: \textbf{(A)}~early functional prototype; \textbf{(B)}~computational material explorations (electronics and textiles); \textbf{(C)}~paper model; \textbf{(D-top)}~refined interim prototype; \textbf{(D-bottom)}~double-layer knit with silver-plated conductive yarn (outer) and polyester (inner).}
  \Description{Teaser collage showing prototype iterations from early electronics on a strap to a knitted band.}
  \label{fig:teaser}
\end{teaserfigure}

\maketitle

\begin{flushleft}
\small
\equalcontrib\ Equal contribution.\quad \equalcorr\ Equal corresponding authors.\\
\insta\ School of Design, The Hong Kong Polytechnic University\\
\instb\ School of Fashion and Textiles, The Hong Kong Polytechnic University\\
\instc\ Department of Language Science and Technology, The Hong Kong Polytechnic University\\
\instd\ School of System Design and Intelligent Manufacturing, Southern University of Science and Technology
\end{flushleft}

\section{Introduction}

E-textile projects often bring together textile designers and interaction designers who approach the same artifact from different perspectives, drawing on expertise that rarely overlaps \cite{posch2019etextilesb,toussaint2020keepinga,hu2023valuecreatinga}. A textile designer reasons about stitch density, layer order, yarn behavior, and machine settings. An interaction designer reasons about signal direction, exposed electrodes, and insulation. Both may be talking about the same part of the same artifact, yet neither description maps onto the other directly. When that translation fails, creative ideas get stuck inside one discipline and never reach the shared design.

\textit{CapSenseBand} was developed to explore that problem through making. The project centered on a knitted capacitive-sensing wristband and sits within a broader line of work on using Swept Frequency Capacitive Sensing to detect microgestures related to stress and emotion \cite{fang2023emosense,fang2024emo}. That program treats wearable sensing as a place where new forms of self-monitoring and expressive wear might take shape, linking the work to creativity for change in how people encounter stress and emotion through cloth and electronics. Our primary interest here was how collaborators could keep creative work moving when the design was still evolving and neither person could fully specify it alone. During the process, Paper Models emerged as a practical communication medium between early material trials and later knitting and electronics decisions, giving both collaborators a way to negotiate orientation, insulation, bodily contact, and assembly without losing momentum.

In this poster, we focus on the design activities: what was tried, what was made, where the collaboration stalled, and how creativity resumed. The artifact matters, but so does the process through which it came into being. For other e-textile teams, the takeaway is a staged rhythm rather than continuous co-design from the start: independent material probes, then convergence through prototypes that each answer a different question: swatches for material feel, a wearable test for bodily mismatch, Paper Models for negotiated form, knit simulation for machine feasibility.

\section{Related Work}

Swatch exchanges, shared material examples, and collaborative prototyping formats help collaborators read material qualities that resist verbal summary \cite{zeagler2013electronic,hertenberger20142013,goveiadarocha2022makinga,jones2020swatch}. Other work has focused on tools that lower barriers between textile practice and interactive technology, including systems for textile patterning, augmented knitting, embroidery, and prototyping across fashion and technology \cite{devendorf2023adacad,albaugh2023augmenteda,wang2023emtex,seyed2019mannequette,seyed2021rethinking}. These systems matter, but they tend to emphasize tool support or handoff rather than how collaborators translate intent during an ongoing project.

At the same time, material-centered HCI has argued that design knowledge often emerges through direct engagement with matter, process, and fabrication \cite{wiberg2018materiality,karana2016tuning}. In e-textiles, this has led to work that begins with yarns, structures, and fabrication methods rather than treating materials as a later implementation detail \cite{zhang2022integratinga,xing2023puffy}. Related projects show how craft practices shape interactive artifacts through machine knitting, wearable sensing, and shape-changing textiles \cite{nabil2019seamless,tsaknaki2019wearableb,albaugh2021engineeringa,luo2021knitui}. Parallel work on modular affective prototyping couples touch-based sensing with dialogue and embodied feedback in configurable companion forms \cite{fang2026emofriends}. Complementing these system-oriented efforts, recent qualitative work examines how users in non-clinical contexts conceptualize AI-supported emotion monitoring and regulation, including concerns around credibility, privacy, and role boundaries \cite{fang2026user}. Prior HCI work has also treated prototypes as more than representations: they focus discussion, make design qualities available for critique, and keep emerging decisions visible during making \cite{lim2008anatomy,innella2017making,bratteteig2012disentanglinga}.

We extend this work with a documented artifact chain for one wristband: how swatches, a wearable test, Paper Models, and knit simulation carried intent while the design was still open. Rather than asking only how to make an interactive textile, we ask how collaborators can carry detailed design intent across domains before fabrication locks those choices, and which intermediary objects made that translation possible in practice.

\section{Method}

This project follows research through design (RtD) \cite{zimmerman2007research}. Rather than separating problem definition from solution making, we let prototype iterations shape both at once. The work was organized with the Analysis-Synthesis-Detailing (A-S-D) method used in material-centered HCI \cite{xing2023puffy,wiberg2018materiality}. This structure suited the collaboration because it allowed early independent exploration and later moments of focused negotiation.

\subsection{Process Overview}
\begin{figure*}[!t]
\centering
\includegraphics[width=\textwidth]{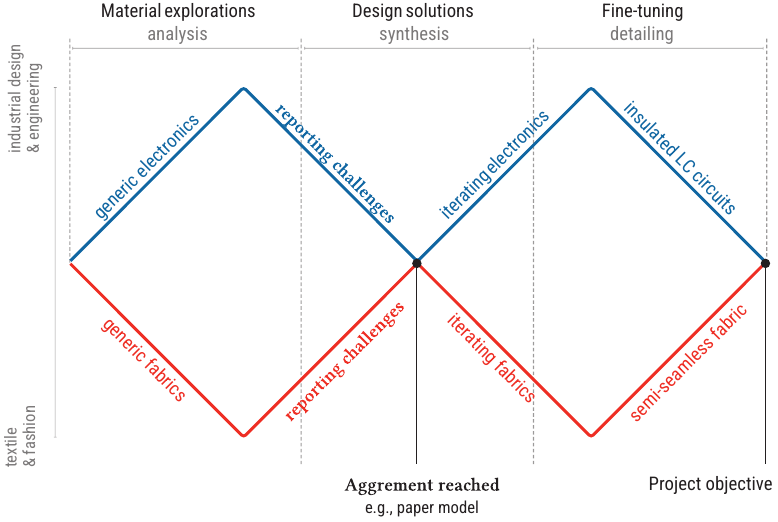}
\caption{The material-centered A-S-D (Analysis-Synthesis-Detailing) design process of \textit{CapSenseBand}. The iterative process moves from independent analysis of electronics and fabrics to collaborative synthesis and finally to detailed refinements, integrating insulated LC circuits with a partly integrated knitted structure.}
\Description{Flow diagram showing the A-S-D process stages: independent analysis of electronics and fabrics, collaborative synthesis into prototypes, and detailing of the final design.}
\label{fig:process}
\end{figure*}

The project began with a shared brief: develop a wrist-worn textile form that could host capacitive sensing while remaining soft, flexible, and close to the body. The brief came from the parent EmoSense program on stress- and emotion-related microgestures \cite{fang2023emosense,fang2024emo}. It fixed the wrist-worn goal and Swept Frequency Capacitive Sensing (SFCS) as the modality for directional, body-close measurement with textile-integrable electrodes; through Analysis, layering, knit structure, openings, and degree of integration stayed open. The brief was functionally narrow and structurally wide: soft, close to the body, capacitive, but not yet specifying how layers, yarns, or integration should be resolved. During analysis, the interaction designer tested sensing electronics and insulation strategies, while the textile designer tested knit structures, yarns, and machine possibilities. During synthesis, both collaborators brought those findings together through physical mockups, Paper Models, and a quick wearable prototype. During detailing, selected ideas were translated into a double-layer knit structure and an insulated SFCS breakout board.

The process moved back and forth more than the diagram suggests. Analysis did not simply end when synthesis began. New samples, machine limits, and board revisions sent the project back into reconsideration. Still, the A-S-D framing is useful because it highlights a shift in collaborative work: early on, each designer mostly asked, ``what can this material do?'' Later, the question became, ``how do we make the same design intent legible to both collaborators?''

\subsection{Team}
The collaboration involved one interaction designer and one textile designer. Each could work quickly within their own tools and materials, but neither could specify the final artifact alone. The interaction designer brought expertise in sensing circuits, board fabrication, and interactive behavior. The textile designer brought expertise in knit structures, yarn selection, garment logic, and flat-bed machine production.

This asymmetry shaped the project. Independent work allowed each collaborator to probe the problem deeply. We kept Analysis separate because joint work too early tended to collapse onto whichever prototype was newest, before knit and circuit constraints were clear \cite{zhang2022integratinga,devendorf2020craftspeoplea}. Shared spatial negotiation waited until each side had material findings worth translating. It also created a translation burden: electrical concerns had to be turned into textile requirements, and textile constraints had to be turned back into implications for sensing and assembly. Shared artifacts became necessary because verbal explanation alone was too imprecise for the level of decision making required.

\subsection{Documentation and Data}
The account presented here draws on project documentation produced during the design work, not on a post-hoc study. Four types of records were retained: physical artifacts (knitted swatches, layered samples, PCB iterations, the final double-layer sleeve); intermediate representations (paper models photographed after each working session, annotated sketches, knit planning files from \textit{SDS-ONE APEX 3}); technical records (schematic and layout revisions, yarn specifications, machine settings, 3D simulation outputs); and reflective notes written by both collaborators after key working sessions. The notes recorded design decisions, points of disagreement, and constraints raised by the machine or circuit that were not otherwise visible in the artifacts. Together these records let us describe design activities with reference to specific artifacts and moments, and trace how the collaboration shifted as the design became more constrained.

\section{Collaborative Design Process}
\label{sec:design}

\subsection{Material Explorations \textit{(Analysis)}}

\begin{figure}[!htb]
\centering
\captionsetup[subfigure]{font=footnotesize,skip=2pt}
\begin{subfigure}[t]{0.28\linewidth}
    \centering
    \includegraphics[width=\linewidth,height=3.2cm,keepaspectratio]{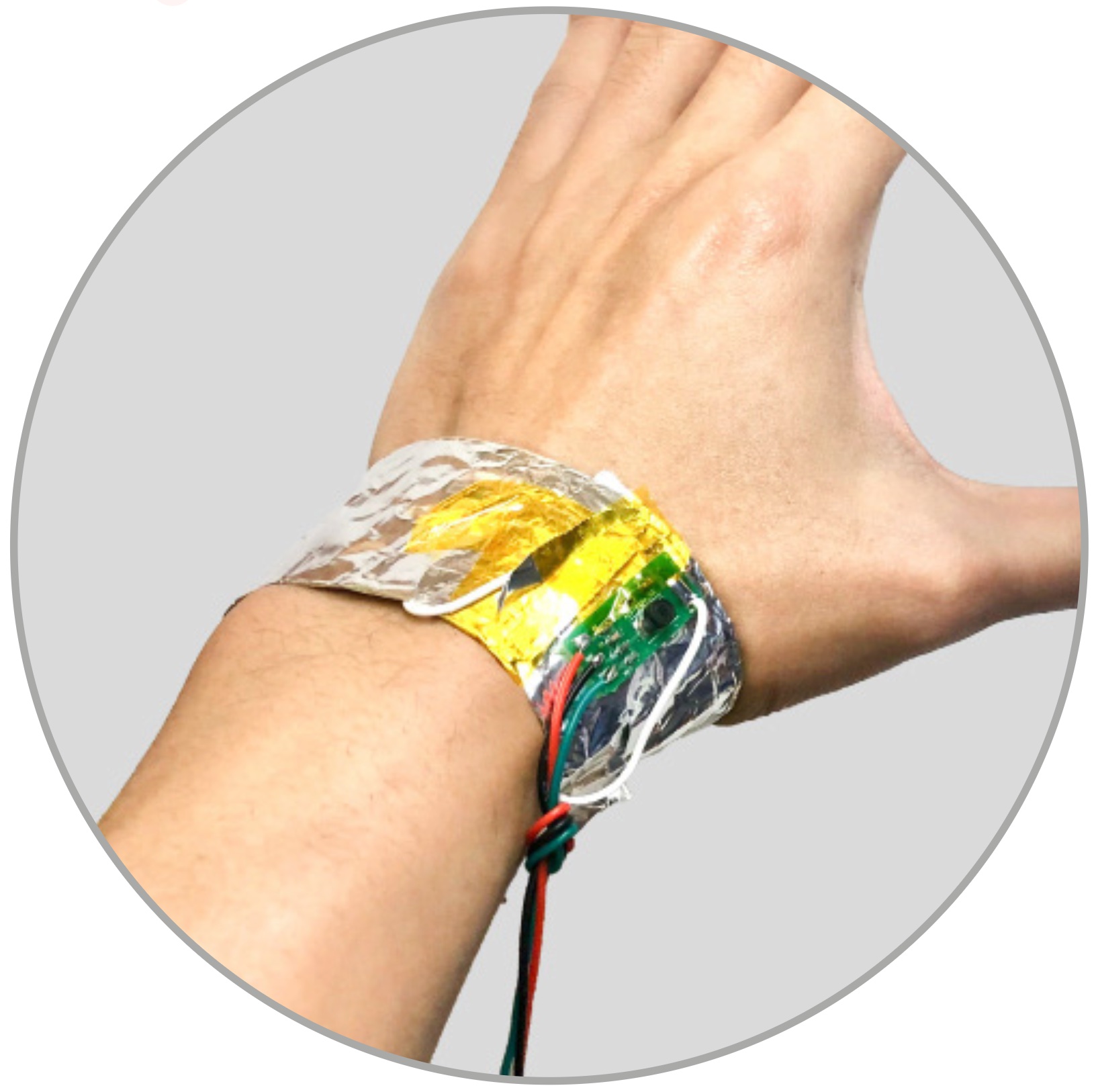}
    \caption[Aluminum foil conductive layer]{Aluminum foil for SFCS testing}
    \label{fig:material-exploration-1}
\end{subfigure}\hfill
\begin{subfigure}[t]{0.31\linewidth}
    \centering
    \includegraphics[width=\linewidth,height=3.2cm,keepaspectratio]{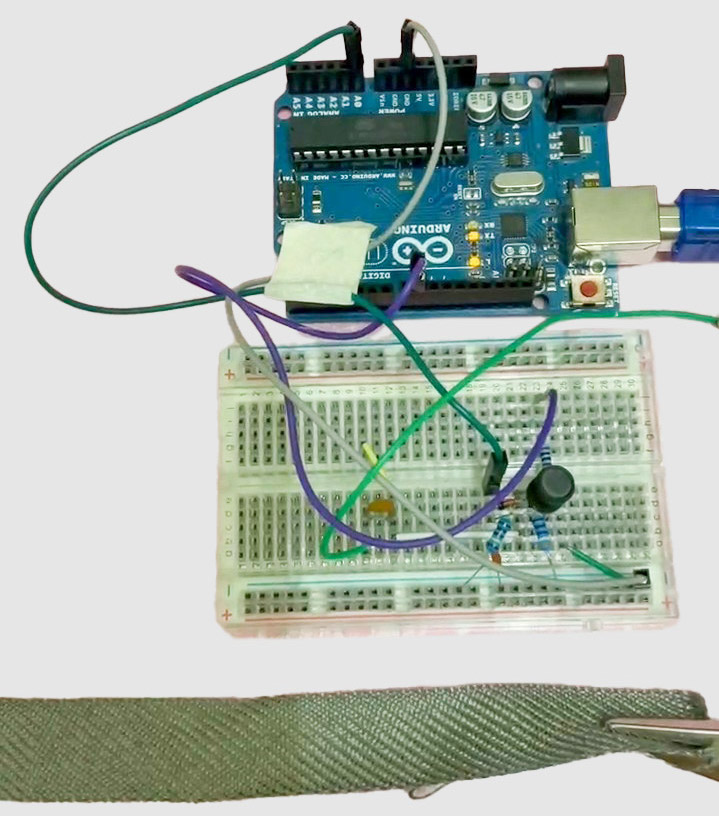}
    \caption[SFCS with Arduino UNO]{SFCS paired with Arduino UNO}
    \label{fig:material-exploration-2}
\end{subfigure}\hfill
\begin{subfigure}[t]{0.31\linewidth}
    \centering
    \includegraphics[width=\linewidth,height=3.2cm,keepaspectratio]{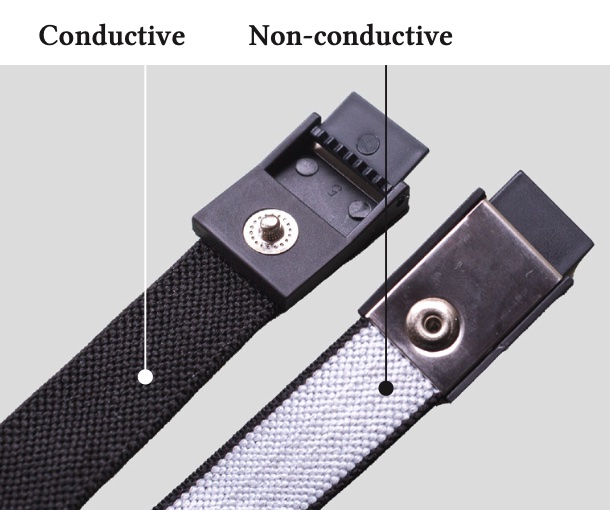}
    \caption[Single-side conductive fabric]{Single-side conductive fabric samples}
    \label{fig:material-exploration-4}
\end{subfigure}
\caption{Material exploration for \textit{CapSenseBand}. Subfigure~(a) uses aluminum foil as a conductive layer for SFCS testing; (b) pairs SFCS with an Arduino UNO; (c) shows generic single-side conductive fabric samples.}
\Description{Three photos showing: aluminum foil used as a conductive layer for SFCS testing, SFCS paired with Arduino UNO, and generic single-side conductive fabric samples.}
\label{fig:material-exploration}
\end{figure}

\textbf{Electronics exploration.} The interaction designer explored Swept Frequency Capacitive Sensing (SFCS) as the sensing approach \cite{sato2012touch,luo2022digital}, chosen to support directional zones close to the body with electrodes that could later be integrated into knit. Early tests used aluminum foil, generic fabrics, and an Arduino-based setup to understand how sensing behavior changed with material arrangement. A basic pattern emerged quickly: the wristband needed directional sensing. The conductive area had to remain responsive, while the skin-facing side had to reduce unwanted signal interference. This was not only a circuit problem. It was also a structural one.

\textbf{Textile exploration.} In parallel, the textile designer explored off-the-shelf wristbands, knitted swatches, conductive yarns, and layered fabric structures. These trials focused on stretch, density, coverage, surface feel, and whether one side of the textile could remain conductive while the skin-facing side stayed insulated. Different stitch types and yarn combinations were tested to understand what kinds of structures might support that functional split.

\textbf{Convergence.} Both lines of work pointed toward a layered structure rather than a single uniform band. The project began to converge on three functional regions: a skin-facing substrate, a conductive layer connected to sensing, and a covering or support layer to maintain orientation. Those regions later reappeared as marked zones on the Paper Model. At this stage, the collaborators shared a design direction but not yet a shared language for specifying it.

\begin{figure}[!htb]
\centering
\captionsetup[subfigure]{font=footnotesize,skip=2pt}
\begin{subfigure}[t]{0.31\linewidth}
    \centering
    \includegraphics[width=\linewidth,height=3.6cm,keepaspectratio]{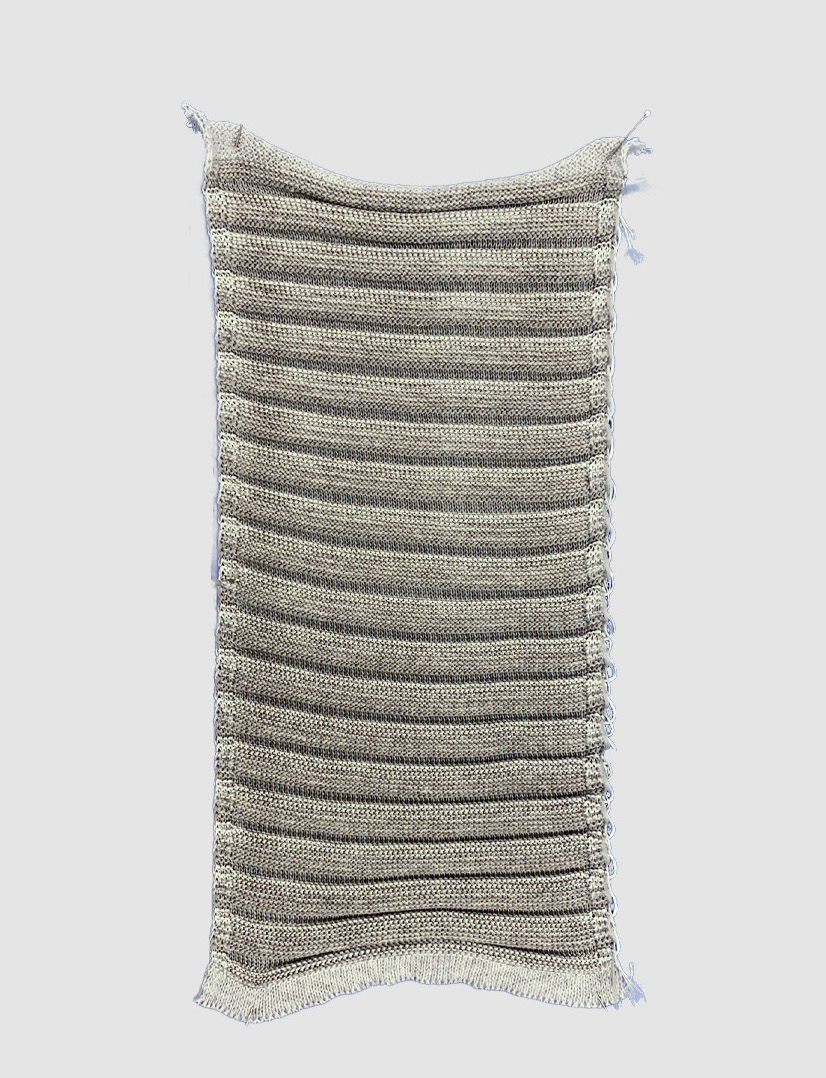}
    \caption[Ribbed fabric sample]{Ribbed Fabric}
    \label{fig:combing-textiles-1}
\end{subfigure}\hfill
\begin{subfigure}[t]{0.31\linewidth}
    \centering
    \includegraphics[width=\linewidth,height=3.6cm,keepaspectratio]{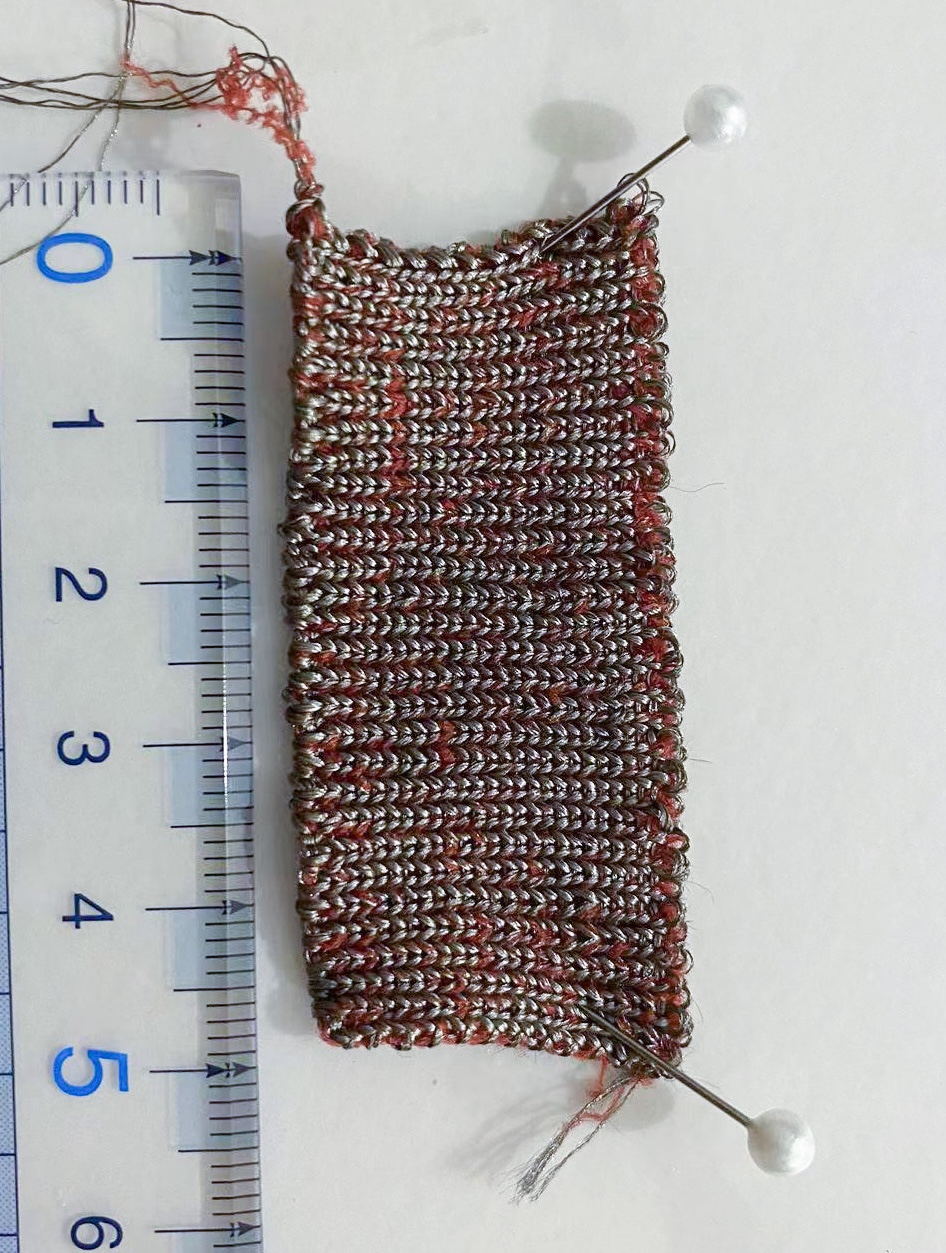}
    \caption[Two-color tuck stitch]{Tuck Stitch Swatch}
    \label{fig:combing-textiles-2}
\end{subfigure}\hfill
\begin{subfigure}[t]{0.31\linewidth}
    \centering
    \includegraphics[width=\linewidth,height=3.6cm,keepaspectratio]{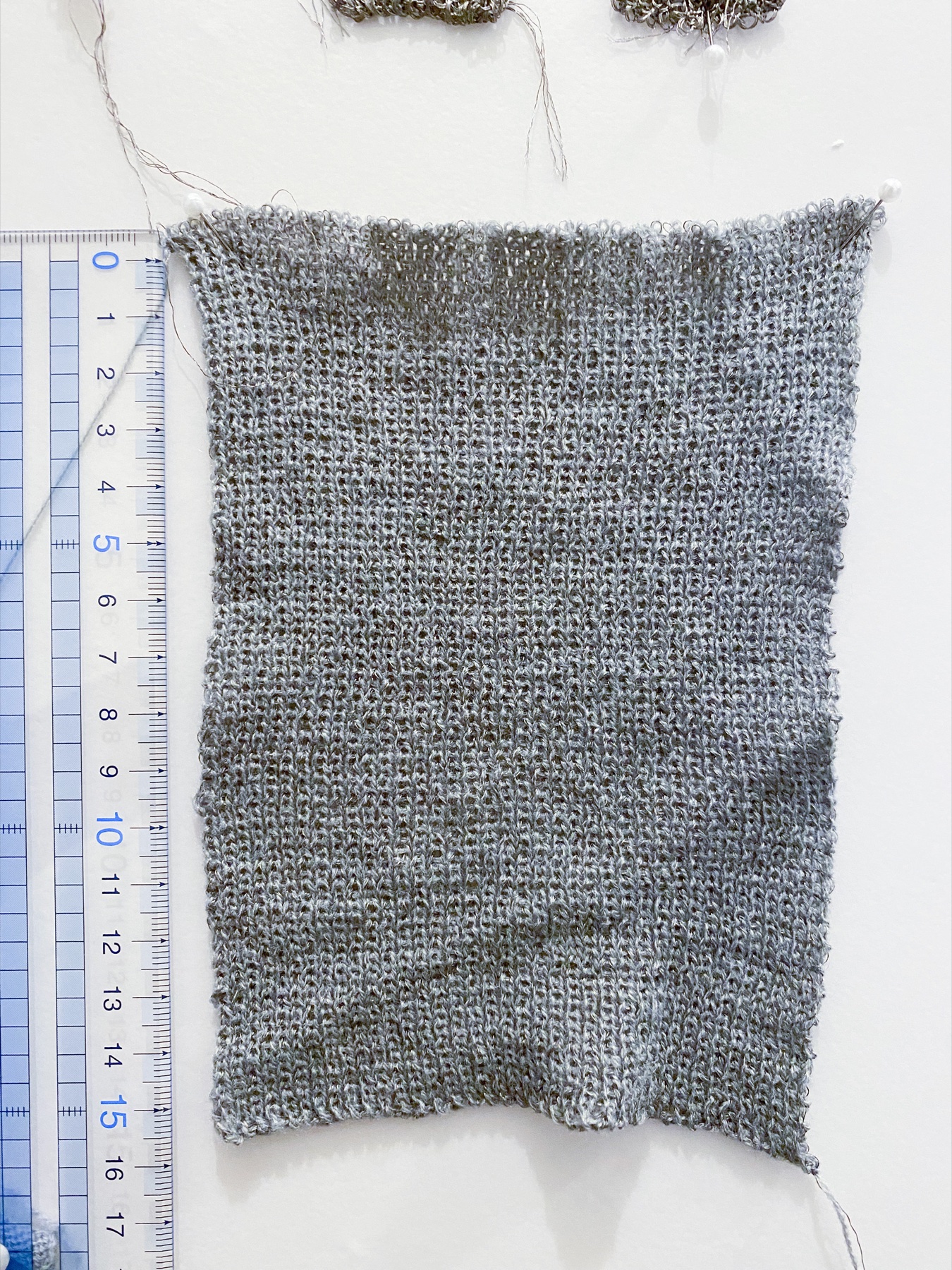}
    \caption[Open-mesh knit swatch]{Open-Mesh Swatch}
    \label{fig:combing-textiles-3}
\end{subfigure}
\caption{Combing and folding textiles exploration. Three distinct machine-knitted fabric samples were explored, from a classic, elastic rib structure to a dense, multicolored tuck stitch and a lightweight, open mesh; these samples represent typical stages in textile development.}
\Description{Three machine-knitted fabric swatches: a ribbed fabric, a two-color tuck stitch swatch, and an open-mesh knit swatch.}
\label{fig:combing-textiles}
\end{figure}

\begin{figure}[!htb]
\centering
\captionsetup[subfigure]{font=footnotesize,skip=2pt}
\newlength{\stockinettepanelheight}
\setlength{\stockinettepanelheight}{3.2cm}
\begin{subfigure}[t]{0.31\linewidth}
    \centering
    \begin{minipage}[c][\stockinettepanelheight][c]{\linewidth}
        \centering
        \includegraphics[height=\stockinettepanelheight,keepaspectratio,trim=780 0 920 0,clip]{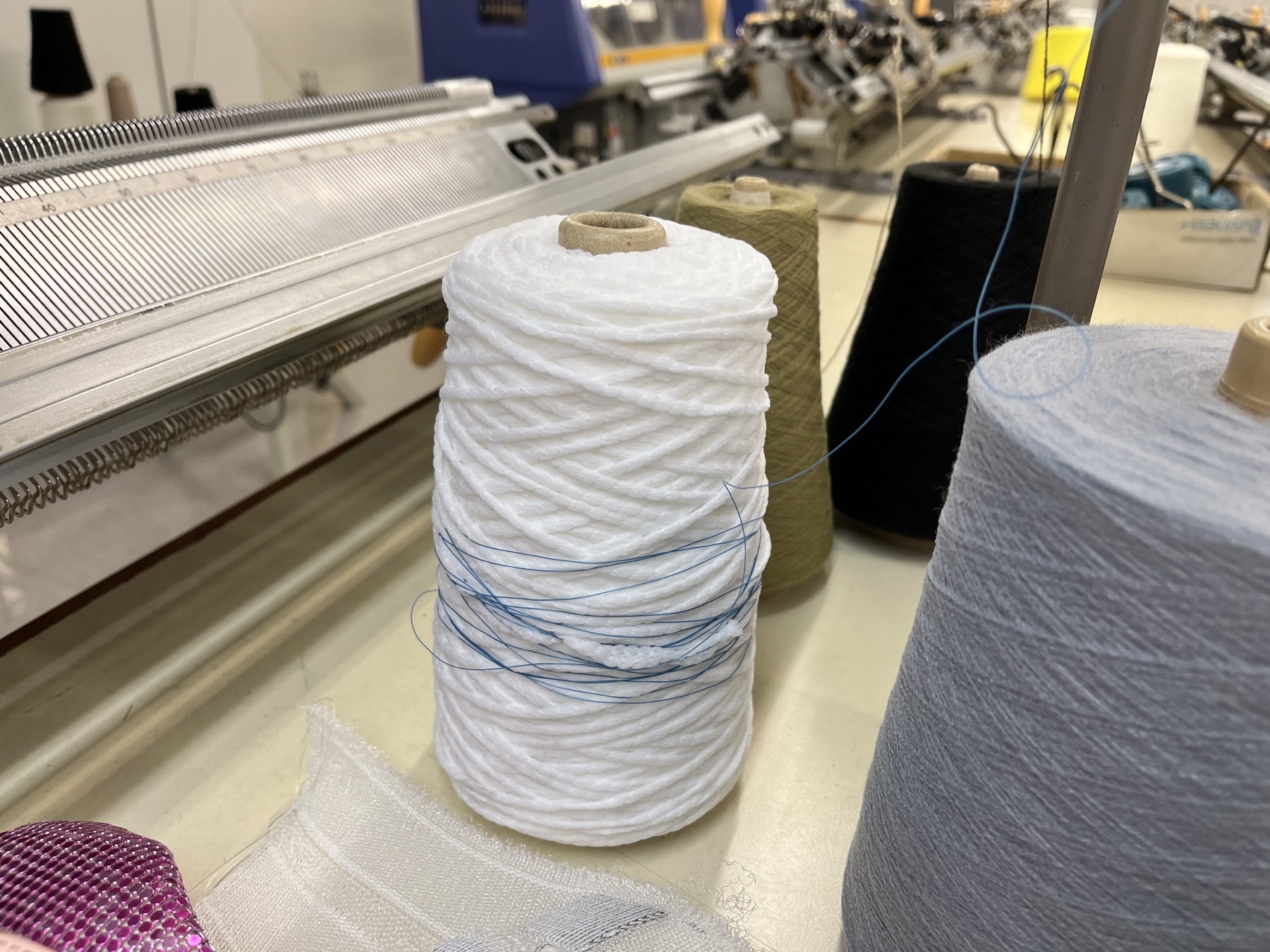}
    \end{minipage}
    \caption[Teflon-coated wires preparation]{Threads with teflon-coated wires}
    \label{fig:stockinette-teflon-1}
\end{subfigure}\hfill
\begin{subfigure}[t]{0.31\linewidth}
    \centering
    \begin{minipage}[c][\stockinettepanelheight][c]{\linewidth}
        \centering
        \rotatebox{-90}{%
            \includegraphics[width=\stockinettepanelheight,keepaspectratio]{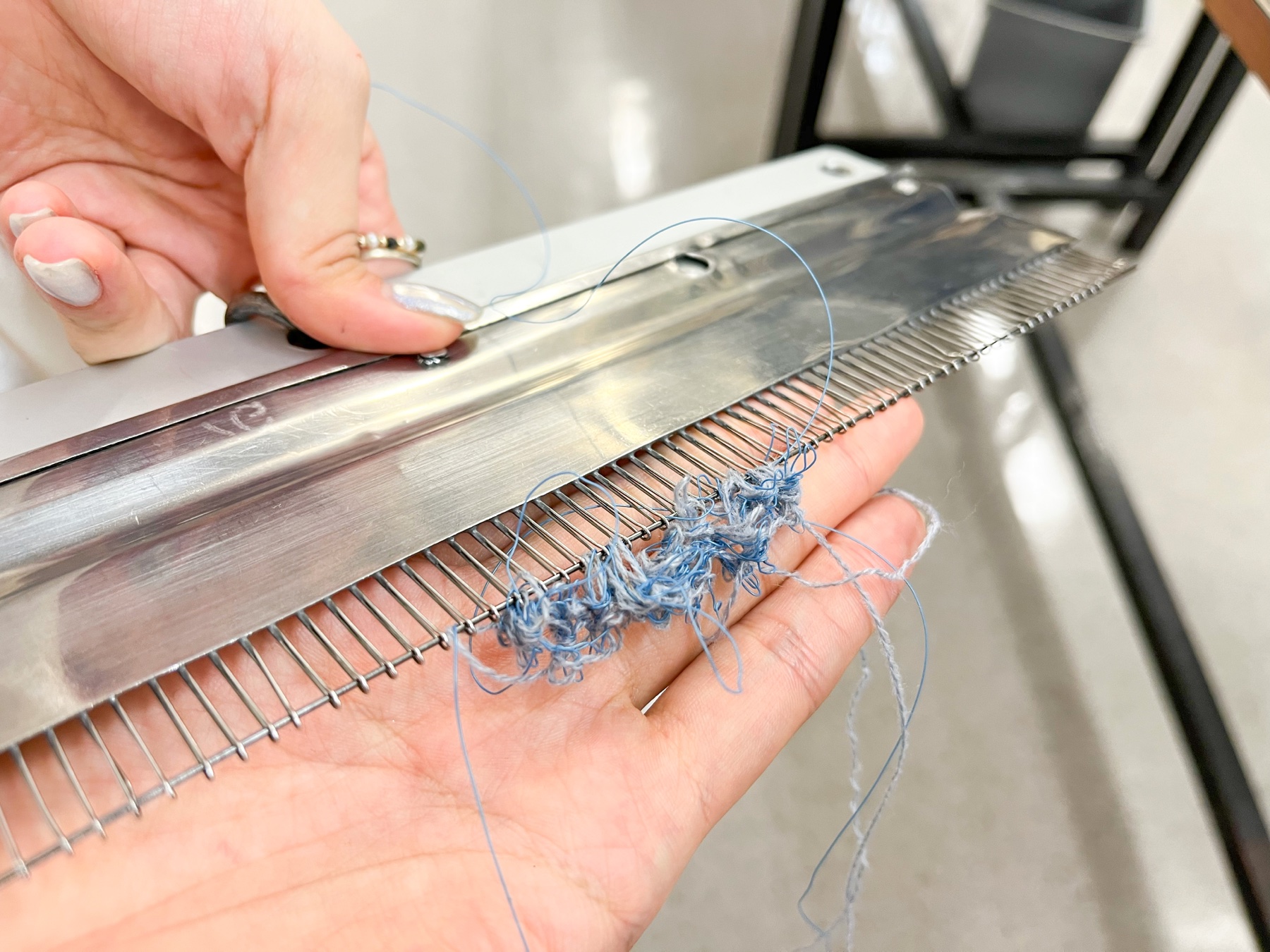}%
        }
    \end{minipage}
    \caption[Stockinette close-up view]{Tangled threads and wires}
    \label{fig:stockinette-teflon-2}
\end{subfigure}\hfill
\begin{subfigure}[t]{0.31\linewidth}
    \centering
    \begin{minipage}[c][\stockinettepanelheight][c]{\linewidth}
        \centering
        \rotatebox{-90}{%
            \includegraphics[width=\stockinettepanelheight,keepaspectratio]{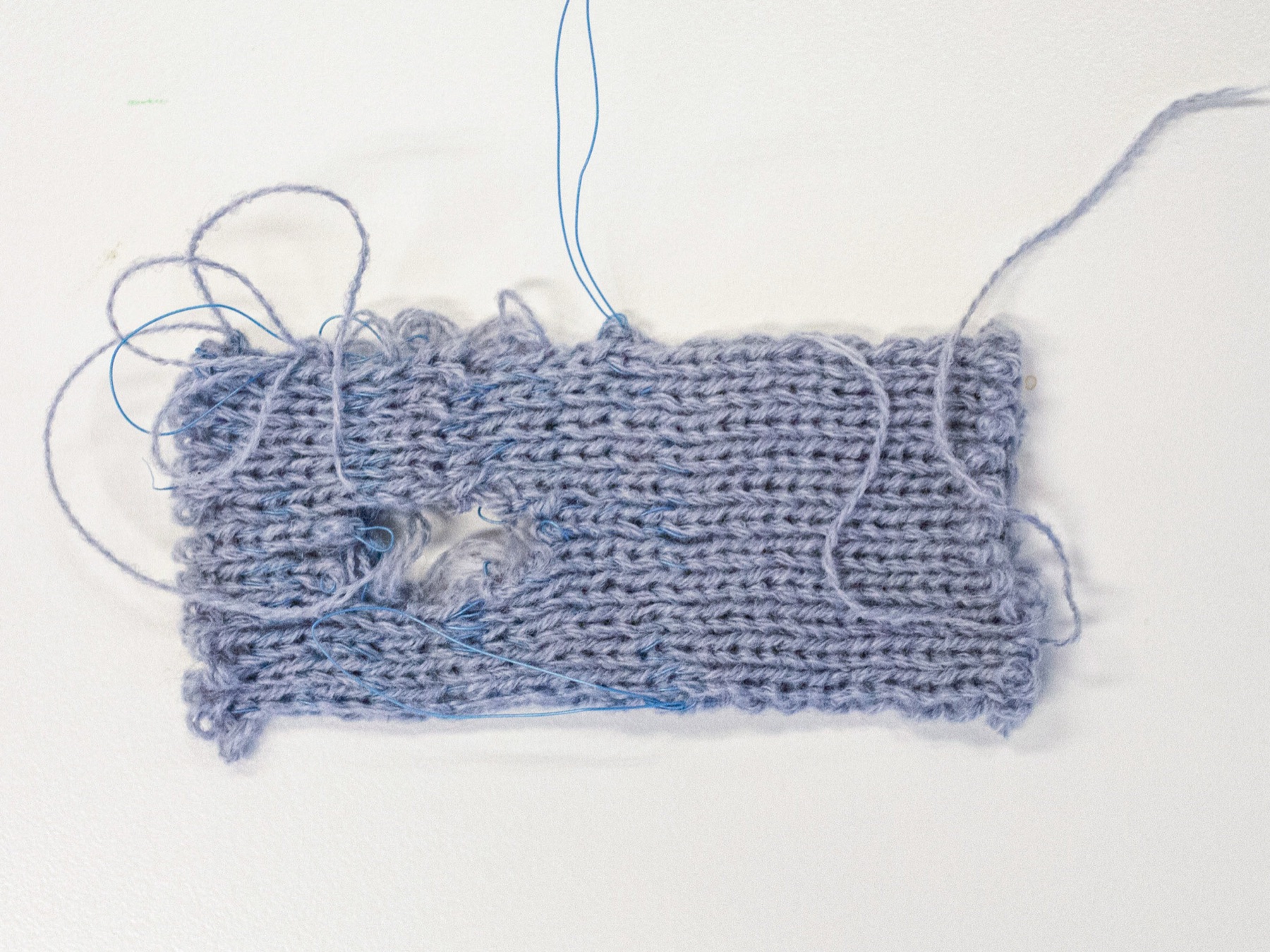}%
        }
    \end{minipage}
    \caption[Stockinette composite result]{Stockinette--wire composite}
    \label{fig:stockinette-teflon-3}
\end{subfigure}
\caption{Stockinette stitch knitted with Teflon-coated wires attempt. Subfigure~(b) shows threads and Teflon-coated wires tangled together, causing failure in the woven surface.}
\Description{Three photos showing the process of knitting stockinette stitch with Teflon-coated wires: thread preparation, close-up of the knit, and the resulting composite.}
\label{fig:stockinette-teflon-attempt}
\end{figure}

The swatch explorations in Figures~\ref{fig:combing-textiles} and \ref{fig:stockinette-teflon-attempt} made otherwise abstract design qualities visible and tangible. Instead of communicating only through general goals, the collaborators could point to a sample and identify what specifically was working or failing. One sample might communicate 'too open, another 'too rigid,' another 'good thickness but wrong contact behavior.' Each swatch answered a discipline-local question that synthesis later had to reconcile.

\begin{figure}[!htb]
\includegraphics[width=\linewidth]{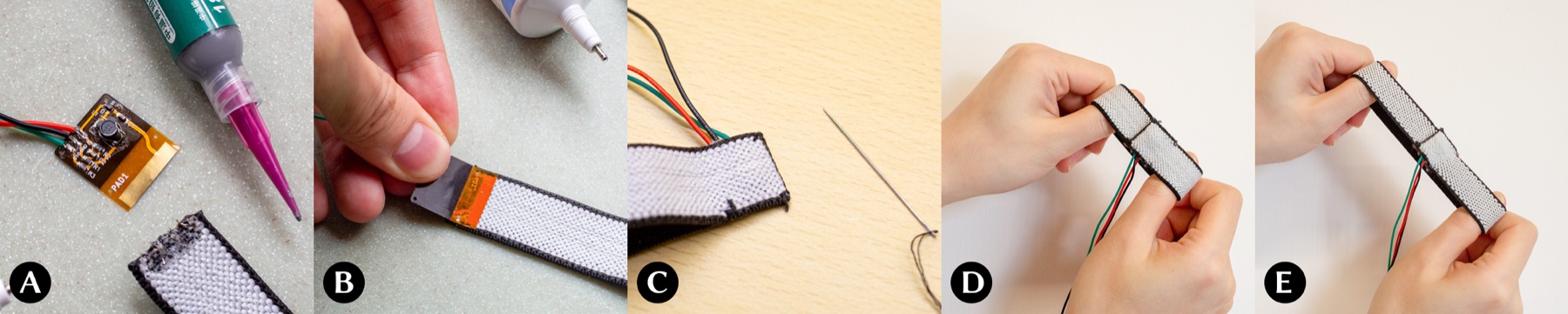}
\caption{\label{fig:quick-and-dirty}The manual fabrication process of \textit{CapSenseBand}. \textbf{(A)} The flexible circuit board extends its capacitive sensing to the one-side conductive band (white) by exposing a golden electrode for connection, leaving the non-conductive side contacting with human skin. \textbf{(B)} Attaching and affixing the circuit board onto the band. \textbf{(C)} Stitching the circuit board onto the band. \textbf{(D)} Assembled band in plain. \textbf{(E)} Assembled band with pulling.}
\Description{Five-panel photo showing the manual assembly of CapSenseBand: PCB electrode connection, attaching and stitching the circuit board to the band, and the assembled band flat and under tension.}
\end{figure}

The quick prototype in Figure~\ref{fig:quick-and-dirty} translated those material findings into a wearable test. By miniaturizing the PCB and attaching it to an off-the-shelf anti-static wristband, the team could move from separated trials to a body-worn proof of concept. The result validated the overall direction but also exposed the primary mismatch: the skin-facing side was not sufficiently insulated, and contact with the body affected sensing in unintended ways.

That failure was productive. It gave both collaborators a shared reference point and made the insulation problem something that could be observed on the wrist and traced to both textile and electronics choices. The wearable prototype forced the collaboration into a more tightly coupled process: a textile change might improve comfort but alter sensing, and a circuit change might improve insulation but complicate assembly.

\subsection{Design Solutions \textit{(Synthesis)}}
\begin{figure}[!htb]
\centering
\captionsetup[subfigure]{font=footnotesize,skip=2pt}
\newlength{\sketchpanelheight}
\setlength{\sketchpanelheight}{3.4cm}
\begin{subfigure}[t]{0.64\linewidth}
    \centering
    \begin{minipage}[c][\sketchpanelheight][c]{\linewidth}
        \centering
        \includegraphics[height=\sketchpanelheight,keepaspectratio]{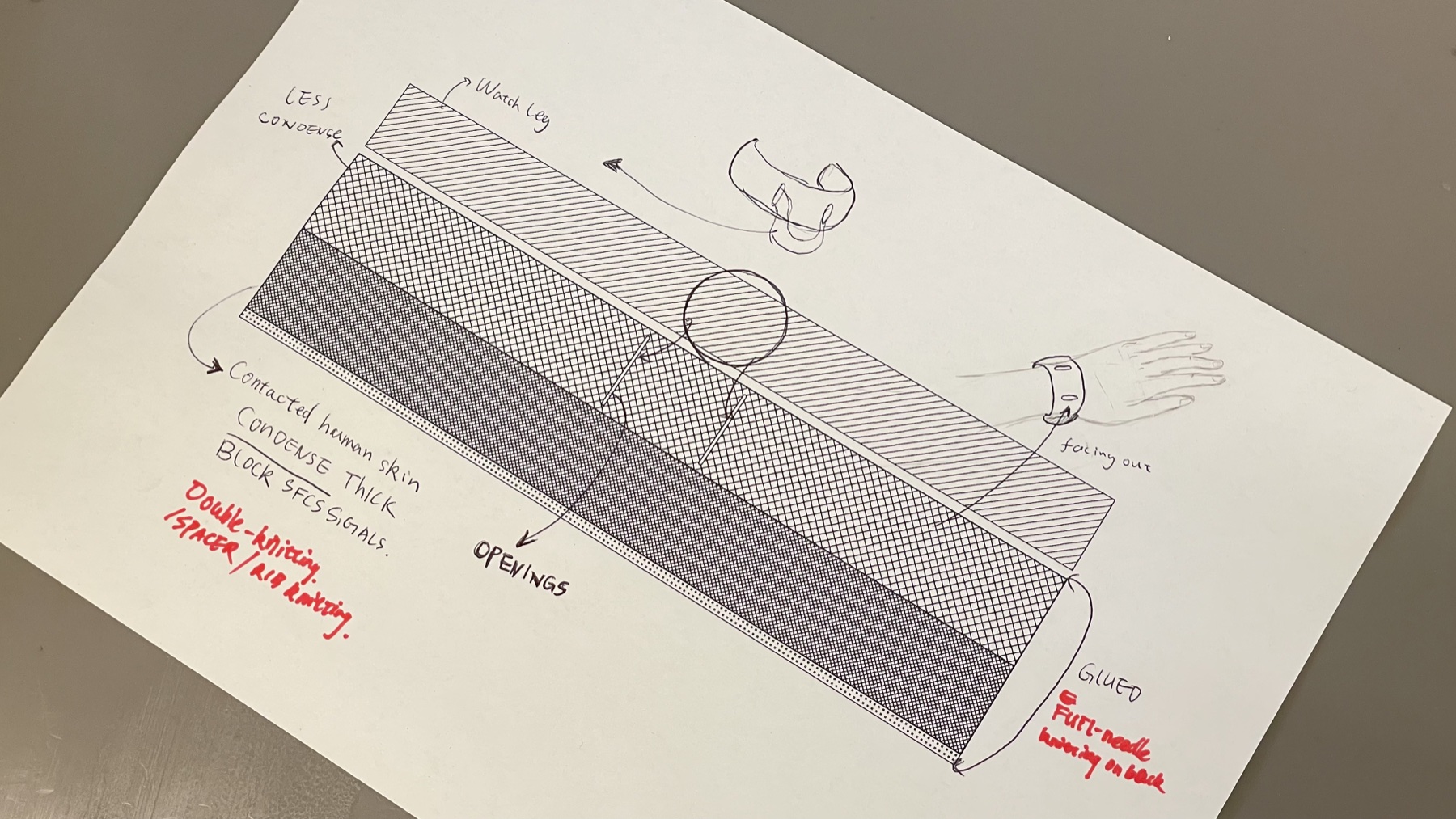}
    \end{minipage}
    \caption[Paper model unfolded]{Unfolded paper model}
    \label{fig:sketch-paper-unfolded}
\end{subfigure}\hfill
\begin{subfigure}[t]{0.32\linewidth}
    \centering
    \begin{minipage}[c][\sketchpanelheight][c]{\linewidth}
        \centering
        \includegraphics[height=\sketchpanelheight,keepaspectratio,trim=520 0 520 0,clip]{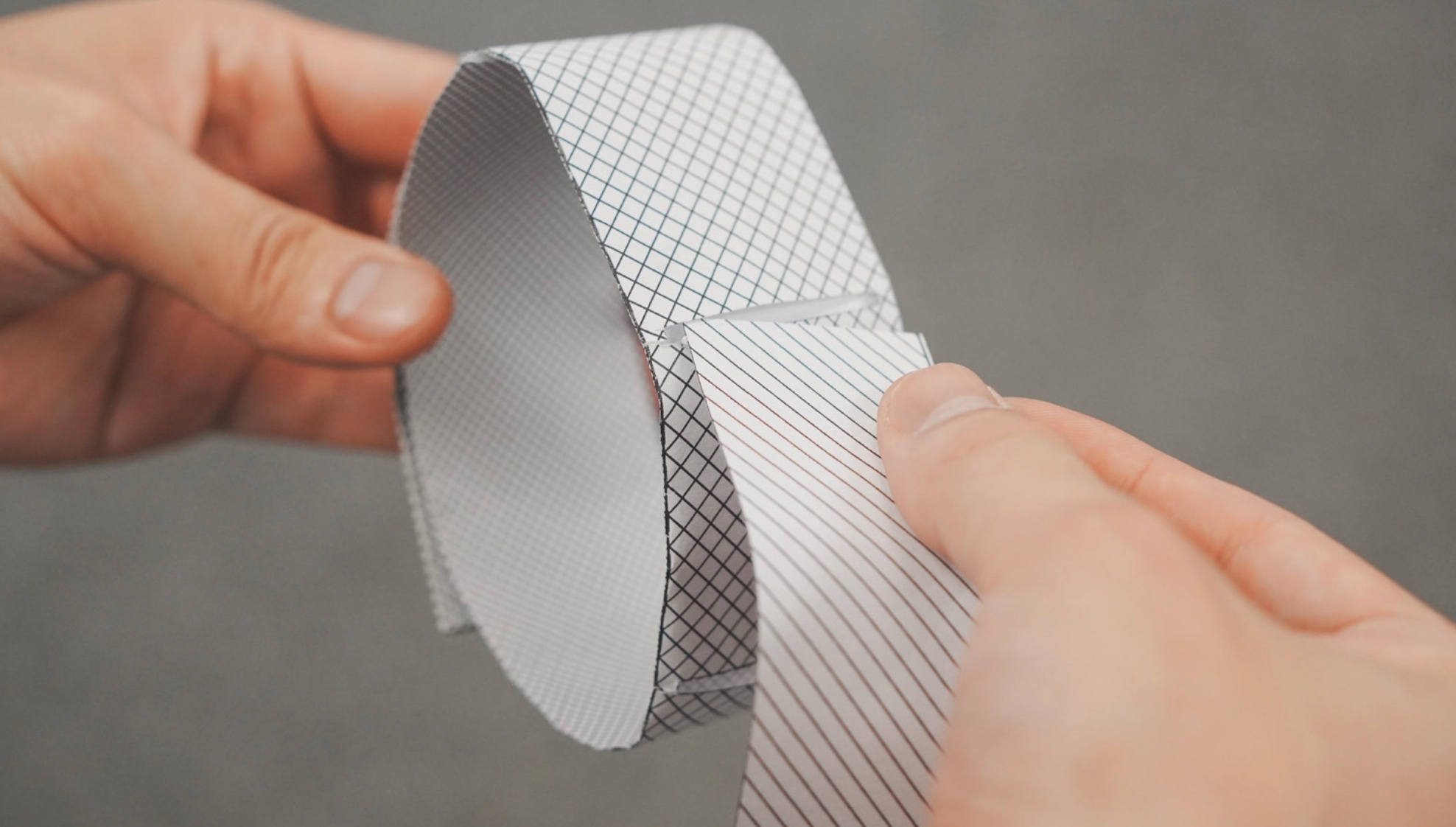}
    \end{minipage}
    \caption[Paper model folded]{Folded paper model}
    \label{fig:sketch-paper-folded}
\end{subfigure}
\caption{Communication with paper model. Subfigure~(a) shows the model flat and unfolded; (b) shows the folded state.}
\Description{Two photos of a paper model used to communicate the wristband design: one in flat unfolded state and one in folded state.}
\label{fig:sketch-paper}
\end{figure}

\paragraph{Paper Models}
The need for a shared proposal led the team to Paper Models. Without planning them as a formal method, the interaction designer began folding and marking paper at the scale of the wristband to show where conductivity, insulation, openings, and body contact should occur. The textile designer annotated the same surfaces with knit structures and fabrication adjustments. Their strength was proximity to the actual wristband form. Both collaborators could point to the same edge, fold, or surface and discuss what it needed to do.

This mattered because the same requirement could take very different forms in each field. The interaction designer might describe a need to block or redirect electrical influence near the skin. The textile designer would need to answer that request through density, layering, gauge, or the relation between different yarns. The Paper Model did not erase this difference. It gave the collaborators a stable object through which translation could happen. Folding the model around a watch body also made a strap passage visible that neither collaborator had named in prior discussions; that fold became the opening specification for knit planning.

The models were especially useful for discussing orientation and assembly. They could show where a watch body or strap should pass through, which side of the structure had to remain conductive, and where openings or folds would matter during knitting. They were quick to revise but specific enough to carry design intent from one meeting to the next.

Paper Models also preserved continuity across meetings. Instead of restarting discussions from memory, the collaborators could return to the same marked object and continue refining it. New annotations accumulated rather than replacing earlier ones. Across three working sessions the same sheet gained layers of markup: first only conductive versus skin-facing zones; then fold lines for strap routing; finally tuck-density notes on the inner face. Photographs kept after each session preserved that sequence in the project archive. In practice, the model became a running record of negotiated decisions about contact zones, attachment logic, material thickness, and the relation between electronics and textile layers.

\subsection{Semi-Seamless Knitted \textit{(Detailing)}}
\begin{figure*}[!t]
\centering
\includegraphics[width=\textwidth]{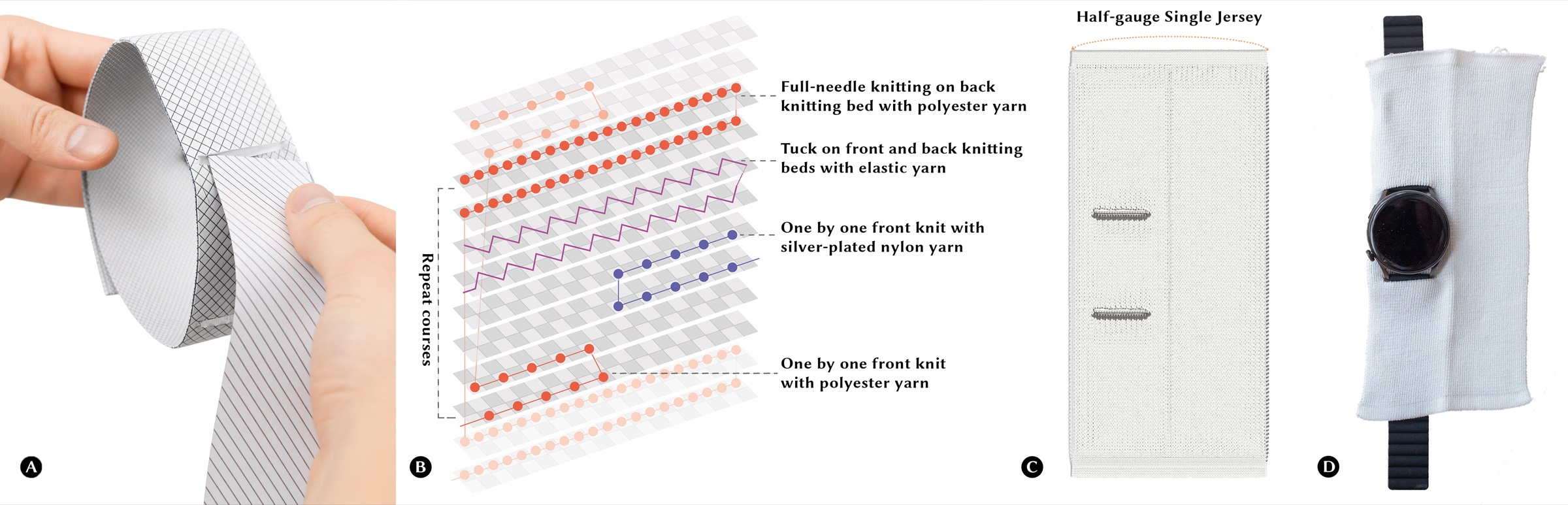}
\caption{Characteristics and technical overview of the double-layer textile for the \textit{CapSenseBand} (panels \textbf{A--D} left to right). \textbf{(A)} Interaction-design sketch of the wristband: conductive outer regions, insulated inner (skin-facing) surface, and openings for watchband legs. \textbf{(B)} Textile designer translated the paper model into knitting actions and yarn choices for the double-layer structure. \textbf{(C)} Knitting quality checked through 3D simulation in \textit{SDS-ONE APEX 3}. \textbf{(D)} Partly integrated double-layer knitted sleeve on a watch.}
\Description{Multi-panel figure showing: interaction designer sketch of wristband characteristics, knitting action diagram, 3D simulation render, and the final partly integrated knitted sleeve on a watch.}
\label{fig:knitted-fabric}
\end{figure*}

The Paper Model was then translated into a double-layer knitted structure. The resulting textile placed a conductive surface where sensing was needed and a denser non-conductive surface against the skin. Knit planning and simulation tested whether this translation could be executed on the available machine before producing the sleeve shown in Figure~\ref{fig:knitted-fabric}.

This translation required several decisions. The outer face combined conductive and non-conductive regions with openings for assembly. The inner face, skin-facing, was knitted to create a smoother and denser insulating surface. Elastic yarn connected the layers so the structure would remain wearable rather than becoming a stiff technical sample. What had begun as a folded paper proposal now had to survive as machine logic, yarn choice, and stitch sequence.

The Paper Model had already aligned the team on double-layer logic and openings, but flat-bed gauge, yarn allocation, and how far the sleeve could be integrated on the machine were not readable from the fold alone. Swatches and simulation (Figure~\ref{fig:knitted-fabric}, panel C) reopened negotiation toward a partly integrated sleeve rather than the fuller integration first sketched.

Machine constraints also reshaped the design. The team first imagined a more fully integrated wristband structure, but available flat-bed knitting conditions made that difficult. The partly integrated sleeve was therefore not only a technical compromise. It was the version of the design that could preserve the most important interaction requirements while remaining fabricable.

This stage made the textile designer's contribution especially visible. The translation from a paper form into knitting action required decisions about gauge, yarn allocation, layer connection, and how openings could be built without destabilizing the overall structure. At the same time, the interaction designer needed to judge whether these textile changes still protected the sensing logic that motivated the design. The detailing phase was less about handing work from one collaborator to the other and more about repeated calibration between the two.

\subsection{Electronics \textit{(Detailing)}}
The board was also revised to match the textile logic. The sensing electrode remained exposed, while the rest of the circuit was insulated with solder mask and silicone rubber. Separating the exposed sensing area from the insulated component side made the board compatible with the textile strategy developed through the Paper Model and the knit structure.

This final stage showed that the project did not end with a single decisive solution. It ended with a coordinated set of adjustments across both domains. The knit sleeve, the exposed sensing region, the insulated PCB body, and the assembly relation between them all had to agree. Together, they formed the current \textit{CapSenseBand} prototype.

Seen across the whole process, the design activities formed an artifact chain rather than a straight line. Material samples informed the quick prototype. The quick prototype exposed the insulation problem. Paper Models turned that problem into a shared design proposal. Knit planning and simulation translated the proposal into a structure ready for the machine. Board insulation then aligned the electronics with the textile logic. Each artifact answered a different question, and each carried part of the next step.

\begin{figure}[!htb]
\centering
\includegraphics[width=\linewidth]{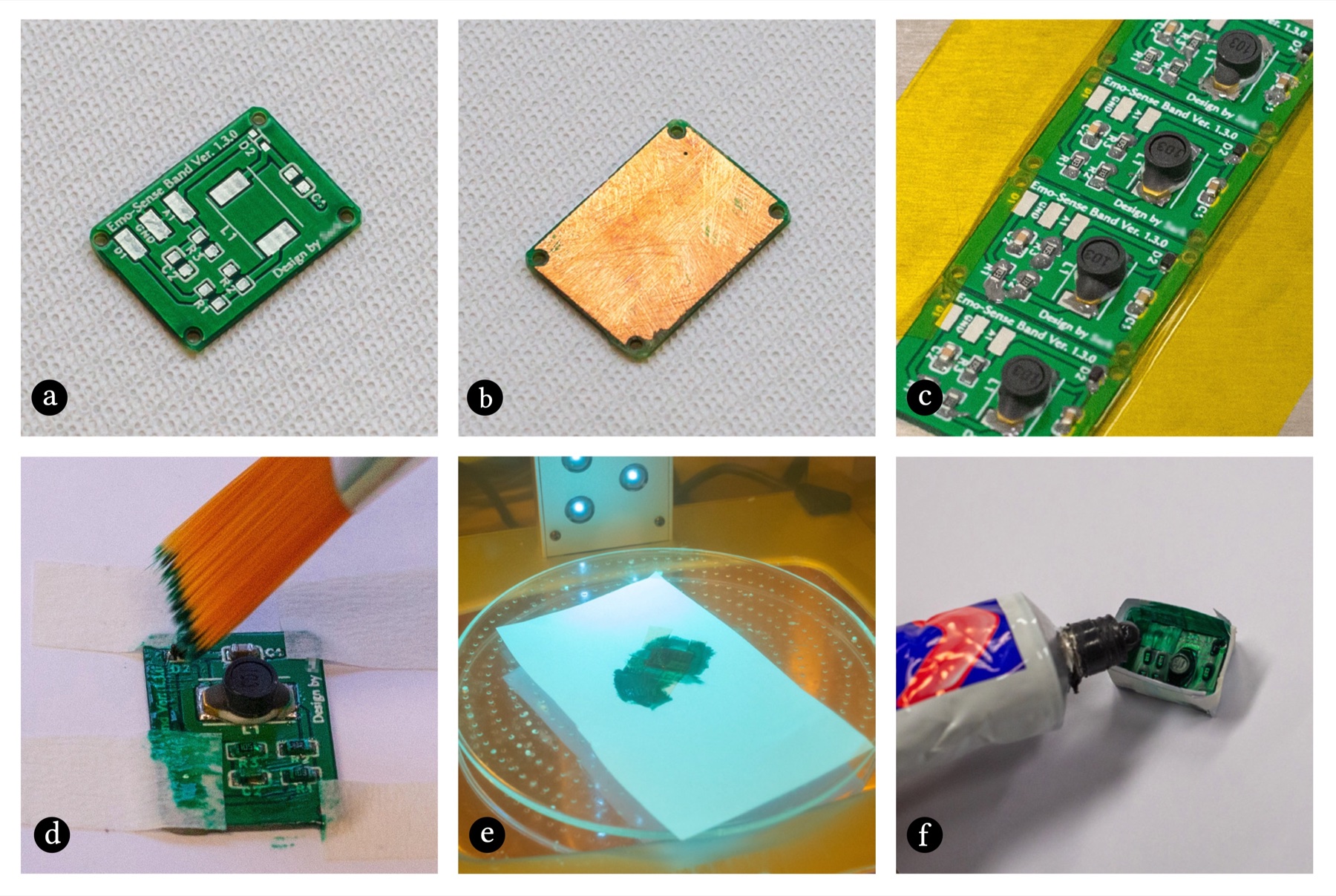}
\caption{Electronic signal insulation process for the SFCS breakout board. SMT components and the sensing electrode are separated across PCB sides; UV solder mask and silicone rubber insulation prevent unwanted AC signal coupling.}
\Description{Photo collage showing six steps (a--f) of the PCB insulation process: component separation, UV solder mask application, and silicone rubber encapsulation.}
\label{fig:electronics-insulation}
\end{figure}

\section{Discussion and Future Work}

\subsection{Paper Models as Boundary Objects}
The Paper Models that emerged in our process resonate with Carlile's account of boundary objects in cross-disciplinary product development \cite{carlile2002pragmatica}. Boundary objects succeed not because they resolve differences between fields, but because they make those differences negotiable at a shared site. Our Paper Models worked in this way. They did not translate textile logic into electronics logic or vice versa. They gave both collaborators a surface on which to register intent, mark disagreements, and accumulate decisions over time. Their value was also generative: because the paper could be folded, marked, and refolded at wristband scale, it produced spatial proposals that neither collaborator had fully specified in words. Lim et al.\ argued that prototypes act as filters, foregrounding some design qualities while setting others aside \cite{lim2008anatomya}. Our Paper Models filtered for exactly the qualities in contest: orientation, layer contact, openings, and insulation boundaries. Material-specific decisions stayed with the person who held that expertise.

This distinguishes Paper Models from other intermediary artifacts in e-textile research. Swatch exchanges carry rich material knowledge but stay anchored in one discipline's vocabulary \cite{hertenberger20142013,goveiadarocha2022makinga}. Prototyping toolkits such as Wearable Bits \cite{jones2020wearablea}, EmTex \cite{wang2023emtex}, and EmoFriends \cite{fang2026emofriends} foreground modular configuration and individual making but do not address cross-disciplinary negotiation while a design is still unsettled. Innella and Rodgers showed that prototypes can make design reasoning visible to others \cite{innella2017making}. Paper Models did something narrower: they gave spatial form to constraints that would otherwise have stayed inside one discipline's description.

\subsection{Alternating Between Independence and Convergence}
Other teams can reuse this rhythm: let each discipline probe materials first, then bring findings to shared objects that make spatial intent legible before fabrication closes options. The design process moved between independent material exploration and joint discussion. Prior work in craft-oriented HCI has found this rhythm productive. Devendorf et al.\ observed that weaving residencies work when craftspeople maintain independent judgment while periodically bringing work to shared evaluation \cite{devendorf2020craftspeoplea}. Zhang et al.\ found that integrating interactive technology concepts with textile expertise requires structured exchange rather than continuous co-design \cite{zhang2022integratinga}. Similar staged translations of intent appear outside e-textile projects: work on embodied pedagogical agents shows how pedagogical intent can move through educator rehearsal, designer interpretation, pose estimation, and agent delivery before being evaluated with students \cite{wei2026grounding}. Our swatches, quick prototypes, and Paper Models each carried different kinds of knowledge. Together, they enabled the team to move forward without forcing either person to fully adopt the other's vocabulary.

Bratteteig and Wagner noted that productive collaboration often depends on preserving, not collapsing, differences in expertise \cite{bratteteig2012disentanglinga}. The collaborative model is part of the contribution here, not only a path to the artifact. Each object in the chain answered a different question: swatches made material properties tangible, the quick prototype located the insulation problem on the body, Paper Models turned that problem into a shared proposal, knit simulation tested whether the proposal was machine-executable. This echoes Buchenau and Suri's point that prototypes derive their value from what they make experienceable at a specific moment in the process \cite{buchenau2000experiencea}.

\subsection{Limitations and Future Work}
Paper Models carried spatial and structural information well, but they did not explain the reasoning behind each annotation. When the interaction designer marked a region as `must remain insulated', the textile designer still needed to understand what insulation means in electrical terms. That conceptual translation happened through conversation, and it was slow. Van Zilt et al.\ encountered a similar gap in multi-disciplinary interactive textile development: domain-specific terminology remained a persistent obstacle even after dedicated tools were in place \cite{vanzilt2022designc}.

Related modular approaches already pair touch-based sensing with LLM-mediated dialogue and embodied feedback for emotion-aware companions \cite{fang2026emofriends}, while user studies in non-clinical settings suggest framing AI as a helper within hybrid workflows rather than as a standalone emotional authority \cite{fang2026user}. Taken together, this suggests one possible response: position a large language model as a listening agent during collaborative sessions. It could generate illustrated explanations of unfamiliar concepts in real time: a capacitive sensing diagram for the textile designer, a tuck stitch illustration for the electronics engineer. It could also annotate shared sketches in parallel layers, one expressing textile implications and another expressing sensing implications, making the translation between domains visible on the artifact itself. A further possibility is speculative bridging: an LLM that carries broad but shallow knowledge across both fields might propose combinations that neither specialist would consider alone, giving both collaborators something concrete to evaluate. None of this would replace Paper Models or material judgment. It would address the problem Paper Models left open: the terminological gap between stitches and signals.

This work is also limited to a single artifact developed by two collaborators. Team composition, institutional context, and available fabrication infrastructure can all shape how knowledge moves across disciplinary boundaries \cite{hu2023valuecreatinga,zeagler2013electronic}. Whether Paper Models remain useful with other wearable forms, or with sensing approaches like textile capacitive and impedance sensing \cite{truong2018capbanda,yu2023uknita}, is not yet known.

Three directions follow from this. We plan to test the collaboration model with other e-textile projects to examine whether the independence-convergence rhythm holds more widely. We want to explore tools that connect sketching, textile planning, and sensing logic more directly, building on existing parametric and augmented design systems \cite{devendorf2023adacad,albaugh2023augmenteda}. And we see room to study how this kind of cross-disciplinary work shapes creativity itself, particularly at moments when an idea must pass through another field before it can be made.

\section{Conclusion}

\textit{CapSenseBand} began as a wrist-worn sensing problem, but the design work soon became a question of how stitches and signals could remain in conversation. Through swatches, quick \& dirty wearable prototypes, Paper Models, knit simulation, and insulated board revisions, we moved between independent disciplinary probing and shared commitment. These activities yielded the final prototype: a semi-seamless integrated knitted sleeve with a conductive outer face, an insulated skin-facing layer, and an SFCS board adjusted to the same textile logic. The process also demonstrates what kinds of objects can sustain cross-disciplinary creativity before a technical textile becomes fixed. Paper Models were useful since they were simple, bodily scaled, and annotatable: they carried orientation, contact, openings, and insulation boundaries without asking either collaborator to abandon disciplinary judgment. For future e-textile work, we suggest treating intermediary artifacts not as rough versions of a final device, but as working surfaces for translation. When collaborators can point to the same fold, stitch region, or sensing boundary, the artifact can keep design intent moving across fields.

\begin{acks}
We thank the UGC Funding Scheme (RHCE \& G.73.xx.R006) at The Hong Kong Polytechnic University for the conference grant and associated funds that helped sustain this project. We also appreciate the research assistants and PhD members at the Research Centre for Future (Caring) Mobility for their constructive prototype suggestions and advice. We thank Dr. Jinyun Zhou from the School of Fashion and Textiles, a sister school within our university, for his digital knitting techniques and guidance in helping produce the prototype samples.
\end{acks}

\bibliographystyle{ACM-Reference-Format}
\bibliography{capsenseband}

\end{document}